
\singlespace
\line{\hfill BROWN-HET-886}
\line{\hfill ANL-HEP-CP-92-118}
\vskip0.25in
\centerline{COLOR--SEXTET QUARK CONDENSATION IN QCD}
\vskip .75in

\centerline{Kyungsik KANG}
\centerline{{\it Department of Physics}}
\centerline{{\it Brown University, Providence, RI  02912} }
\vskip .10in

\centerline{and}
\vskip .10in

\centerline{Alan R. WHITE}
\centerline{{\it High Energy Physics Division}}
\centerline{{\it Argonne National Laboratory, Argonne, IL  60439}}
\vskip .25in

\centerline{ABSTRACT}
\vskip .25in

\parindent 40pt
{\narrower\smallskip\noindent{\tenrm We suggest that the sextet color
condensation model is not only the simplest and most attractive scheme
to modify the standard model but also may already have some
``experimental" support.  The crucial test may be to observe a short--
lived axion--like $\eta_6$ of mass around 60 GeV/$c^2$ that can
produce high energy photon pairs diffractively at hadron colliders and
also radiatively from $Z^0$ at high energy $e^+ e^-$ colliders.  Here we
show how high--energy forward scattering can be affected by
diffractive production of the $\eta_6$.  Presented to the DPF-92
meeting.\smallskip}}

\parindent=0pt

\vskip .25in

The color--sextet quark condensation model has a number of attractive
features.  First of all, the chiral symmetry breaking in the color--sextet
quark sector of QCD provides the electroweak symmetry breaking scale
and can lead to many new phenomena due to mixing of electroweak and
strong interactions.  Secondly, it is possible to construct minimal
condensation model $^1$ with a flavor doublet of color--sextet quarks
that has an
infrared fixed point in the beta function, possessing a desired
property of a walking technicolor model.  Thirdly there is a natural
resolution of the strong CP problem in this model involving a special
axion -- the $\eta_6$, which is anomalously heavy because of large
anomalous dimensions of sextet quark operators and multiple instanton
interactions.  We initially $^2$ suggested that the behavior of the
forward elastic $p\bar{p}$ scattering
amplitude measured at CERN SPS and Fermilab Tevatron might be due to
diffractive production of the $\eta_6$ originating from
the sextet--quark condensation.  In particular we pointed out the
``mini--Centauro" and ``Geminion" events in very high--energy cosmic
ray experiments$^3$ could actually be identified as the hadronic and
two--photon decay products of the $\eta_6$ which should be produced
diffractively in hadronic process.  The most
distinctive of the Geminion events $^4$ now seems to prefer a higher
$\eta_6$ mass of 60 GeV.  Also the $\eta_6$ would eventually$^5$ be
observable as a rare radiative decay of $Z^0$ at LEP.  We pointed out
that $^6$ the massive 2$\gamma$ events $^7$ seen by L3 and DELPHI might be
actually due to $\eta_6$ production and that the measurements of R
around 59 GeV at TRISTAN hinted rather large hadronic cross sections.

The minimal sextet--quark condensation model (MSQCM) $^2$ is a
straight--forward modification of the standard SU(3) $\times$
SU(2)$\times$ U(1) model by adding a flavor doublet of color sextet quarks
$Q_6 =$ (U,D)
with $(6^*, 2, 1/3)$ to the usual fermions of the standard model but
with no elementary Higgs scalars.  As the QCD coupling $\alpha_s
(\mu)$ evolves, a $\bar{Q}_6 Q_6$ condensate forms at a high
scale $F_{\pi_6} = \vert \langle \bar{Q}_6 Q_6 \rangle \vert^{1/3}
\sim $ 250 GeV due to ``Casimir rule" producing four massless Goldstone
bosons i.e. an isotriplet
$\pi_6$ and an isosinglet $\eta_6$ of which $\pi_6$ are absorbed by $W^{\pm}$
and $Z^0$ giving the mass relation $M_W = M_Z \cos \theta_{\omega} \sim
gF_{\pi_6}$.  The $\eta_6$ is an axion associated with a U(1) axial chiral
symmetry of the color--sextet
quarks and is expected to be much heavier than the ordinary axion
because of a large
anomalous dimension for sextet quarks operators and multiple
instanton interactions.  The existence of the $\eta_6$ provides a unique
signal of the sextet quark
condensation model, which is unparalleled by other models.
The new QCD baryons and mesons made of sextet
quarks are all expected to be much heavier than $F_{\pi_6}$.

We have argued that$^2$ the behavior of high energy elastic
$p\bar{p}$ scattering at CERN and Fermilab collider energies (i.e.,
the surprisingly ``large" real part of the forward amplitude observed
by UA4 at CERN and rather ``low" total cross-sections measured by E710
and CDF at Fermilab), might be due to diffractive production of the
$\eta_6$ that has the properties of two particular classes of cosmic
events $^3$ known as ``Geminions" or ``binocular" events and
``mini--Centauros".  These cosmic exotics are known to be produced {\bf
diffractively} with a threshold energy around 400-500 GeV and
have a common ``fireball" mass of 60 GeV. The mini--Centauros are hadron
rich with an average
multiplicity 15, while the Geminions have two well--separated showers,
and they both have a cross section of a few percent of the total
cross--sections, thus giving an order of magnitude estimate
$M_{\eta_6}\simeq$ 60 GeV.  If the $\eta_6$ is produced
diffractively in $p\bar{p}$ scattering, a major new threshold of
400--500 GeV in strong interaction is expected to follow.  This
can be the cause for the large UA4 real part, i.e.,
$\rho= ReA/Im A\simeq 0.24$, for the
forward $p\bar{p}$ scattering amplitude $A$.  As long as $\rho$ is
significantly larger than 0.13, the standard value consistent with
smooth and slow rise of the total cross sections, no models $^8$ without
a threshold term
can give satisfactory fit to high--energy forward scattering data.
Conventional models
containing either ln$^2$s or lns Pomeron and Regge terms give
$\rho = 0.13$ at 546
GeV with $\chi^2 $/D.F. = 1.44, when we$^9$ fit 109
data above 9.78 GeV by the Minuit program (See Fig.1a and Fig.1b). The
maximal Odderon model does not fit the data either giving $^9$ $\rho =
1.42 (1.85)$ at 546 GeV (1.8 TeV) (and even larger
$\rho$ beyond 1.8 TeV) and a comparable $\chi^2$/D.F. but with
unstable Regge parameters for the same data set.  The
threshold model containing lns Pomeron and exchange non--degenerate
Regge terms can easily fit the data, giving $\rho = 0.15 - 0.20$ at
546 GeV with a
threshold energy close but below the UA4 energy and similar
$\chi^2$/D.F..
Figs. 2a and 2b show a typical fit by a threshold model with
threshold at 546 GeV, threshold power $\alpha = 1/2$ and
$\chi^2$/D.F. = 1.41.  We stress that the diffractive threshold
picture involving production of the $\eta_6$ is supported by the
existing experiments and will remain to be viable if
$\rho$ is in the range of 0.15 - 0.20.
\vskip6.0in

We may say that the MSQCM as a dynamical symmetry breaking scheme has
numerous virtues and can give rise to a heavy axion $\eta_6$ that can be
produced
diffractively at high--energy hadron colliders and radiated from
$Z^0$ at $e^+ e^-$ colliders.  The $\eta_6$ poses as a strong candidate
for the new particle of mass 0(60 GeV) responsible for the massive photon
pairs reported by L3 and
DELPHI at LEP and has possibly been seen already at Tristan.  A
large real part of the forward scattering amplitude can follow from
diffractive production of the $\eta_6$. Crucial test for the
color--sextet quark sector is then to look for diffractive production of the
$\eta_6$ at the hadron collider with a relatively large
cross--section.  The $\eta_6$ with a mass of 60 GeV was not ruled out
by the preliminary CDF search$^{10}$ and the cross section for the two
photon decay mode is apparently below the
sensitivity of the search.  However the $\eta_6$ could be observed at
the high--energy hadron collider via its decay into a photon pair or a
lepton pair.

Besides the CDF attempt to search for the
$\eta_6$ via $2\gamma$ decay in diffractive $p\bar{p}$ scattering, an
interesting physics possibility$^{11}$ at a photon linear collider has
been proposed to search for a host of new particles including the
$\eta_6$.  In the meantime we are eagerly waiting for the result from
the UA4.2 group$^{12}$ on the real part of the forward $p\bar{p}$
scattering amplitude.
\vskip .25in
\noindent{\bf References}
\vskip .25in

\pointbegin
W. J. Marciano, {\it Phys. Rev.} {\bf D21}, 2425 (1980); E. Braaten,
A. R. White and C. R. Willcox, {\it J. Mod. Phys.} {\bf A1}, 693
(1986); K. Kang and A. R. White, {\it J. Mod. Phys.} {\bf A2}, 409
(1987).
\point
K. Kang and A. R. White, {\it Phys. Rev.} {\bf D42}, 2425 (1989); {\it
Phys. Lett.} {\bf B266}, 147 (1991).  See also (BROWN-HET-877/ANL-HEP-
CP-92-84), Proc. 19th Int. Colloq. GTMP,
Salamanca, Spain (1992) for further references.
\point
S. Hasegawa, in VI International symposium on Very High Energy Cosmic
Ray Interactions, Taubes, France (1990).
\point
T. Arizawa (and also S. Hasegawa), Private Communication.
\point
T. Hatsuda and M. Umezawa, {\it Phys. Lett.} {\bf B254}, 493 (1991).
\point
K. Kang, I. G. Knowles and A. R. White, ANL-HEP-PR-92-66/BROWN-
HET-872; see also A. R. White and K. Kang in this Proc.
\point
The L3 Collab., CERN-PPE/92-152. See also S.S. Ting; B. Wyslouch; and
J. Marco in this Proc.
\point
K. Kang, {\it Nucl. Phys.} {\bf B} (Proc. Suppl.) {\bf 12}, 64 (1990); M. M.
Block, K. Kang and A. R. White, {\it J. Mod. Phys.} {\bf A7}, 4449
(1992); K. Kang and A. R. White, {\it Nucl. Phys.} {\bf B} (Proc.
Suppl.) {\bf 25}, 70 (1992).
\point
K. Kang, P. Valin and A. R. White, BROWN/ANL preprint - to appear.
\point
N. D. Giokaris et al, {\it Nucl. Phys. B} (Proc. Suppl.) {\bf 25B}, 40
(1992).
\point
D. L. Borden, D. Bauer and D. O. Caldwell, SLAC-Pub-5715.
\point
UA4.2 (Genoa-Palaiseau-Praha-Roma-Valencia) Collab.
\end